# Improving the Economic Complexity Index

Saleh Albeaik[1], Mary Kaltenberg[2,3], Mansour Alsaleh[1], Cesar A. Hidalgo[2]

[1] Center for Complex Engineering Systems, King Abdulaziz City for Science and Technology

[2] Collective Learning Group, The MIT Media Lab, Massachusetts Institute of Technology

[3] UNU-MERIT, Maastricht University

**Abstract:**

How much knowledge is there in an economy? In recent years, data on the mix of products that countries export has been used to construct measures of economic complexity that estimate the knowledge available in an economy and predict future economic growth. Here we introduce a new and simpler metric of economic complexity (ECI+) that measures the total exports of an economy corrected by how difficult it is to export each product. We use data from 1973 to 2013 to compare the ability of ECI+, the Economic Complexity Index (ECI), and Fitness complexity, to predict future economic growth using 5, 10, and 20-year panels in a pooled OLS, a random effects model, and a fixed effects model. We find that ECI+ outperforms ECI and Fitness in its ability to predict economic growth and in the consistency of its estimators across most econometric specifications. On average, one standard deviation increase in ECI+ is associated with an increase in annualized growth of about 4% to 5%. We then combine ECI+ with measures of physical capital, human capital, and institutions, to find a robust model of economic growth. The ability of ECI+ to predict growth, and the value of its coefficient, is robust to these controls. Also, we find that human capital, political stability, and control of corruption; are positively associated with future economic growth, and that initial income is negatively associated with growth, in agreement with the traditional growth literature. Finally, we use ECI+ to generate economic growth predictions for the next 20 years and compare these predictions with the ones obtained using ECI and Fitness. These findings improve the methods available to estimate the knowledge intensity of economies and predict future economic growth.

**Introduction**

For decades, the theory and empirics of economic growth has attempted to understand why some economies grow faster than others. The early literature focused on the accumulation of simple productive factors such as labor and physical capital [1]. But soon enough, the literature turned into more nuanced factors, such as human capital [2,3], institutions [4,5], social capital [6–8], and technological change [9,10]. Yet, even when taken together, these factors have been unable to fully explain economic growth. As a result, economic growth still poses questions, embodied in the idea of Total Factor Productivity (TFP), a measure of the output of an economy that is not explained by the availability of factors. That is, a measure of how much output an economy can produce per unit of input.

In the last decade, the search to understand TFP gave rise to a new literature on economic complexity [11–21], which does not aim to identify individual factors, but to measure combinations of them indirectly. The assumption is that, if growth depends on having combinations of factors, and on the ability to use them productively, then, it should be possible to measure the combinations of factors that predict growth—whatever these may be—by looking at the expression of these factors in the diversity and sophistication of the products that countries produce and export.

Consider exporting fresh fish. Fresh fish is a product that requires specific physical capital inputs, such as a reliable power grid and cold storage, but also, that requires specific institutional factors, such as navigating international phytosanitary standards. Producing and exporting fresh fish, however, also requires specific knowledge on aquiculture and on the global fish market. This means that an observation as simple as seeing a country export fresh fish can tell us about the presence of specific technological, human, and institutional factors, in an economy.

Measures of economic complexity have been validated by studying their ability to predict future economic growth. Economic complexity is highly predictive of future economic growth once we control for a country's initial level of income [11,22], and this observation is robust to controlling for a large number of factors, from human capital factors, to measures of competitiveness and institutions [22]. That is, countries with an income that is below what we expect based on its productive structure grow faster than those with an income that is too high. Moreover, recent work has also shown that countries with relatively high levels of economic complexity tend to have lower levels of income inequality, even after controlling for measures of education, income, and institutions [14].

The ability of economic complexity to predict growth supports well-established ideas in economics, such as the idea that institutions, education, knowledge, know-how, and technology, are required for economic growth. What these measures of complexity do differently, however, is that they avoid the need to define these factors or their importance a-priori. Instead, the measures are based on techniques that help define the knowledge intensity of economies and of activities endogenously from the data. These endogenous definitions rely on simple linear algebra techniques. For instance, the original economic complexity index (ECI) [11] defines the complexity of an economy as the average complexity of its products, and the complexity of a product as the average complexity of the countries exporting it. This circular argument is in fact mathematically tractable using linear algebra and has a solution in the form of an eigenvector, allowing the creation of an endogenous definition of the complexity, or knowledge intensity, of an economy.

This technical innovation helped separate these measures of economic complexity from other measures relying on exogenous definitions of knowledge intense activities (efforts, for instance, that define some sectors—e.g. services, or software—as knowledge intense, and then measure knowledge intensity as the fraction of people employed in these sectors). This innovation also helped these measures become adopted in other domains; for instance, they have been used to estimate the innovative capacity of cities using patent data [23].

Yet, these measures of complexity are not free of limitations. One limitation of the economic complexity index is that it requires defining which countries export which products, a task that is not easy to do in a world where the markets for products and the sizes of economies vary by multiple orders of magnitude. The convention has been to consider as exports only the products that a country has a revealed comparative advantage in [24]. Yet, this definition introduces a hard threshold that introduces noise around the boundary. The metric of economic complexity we introduce here (ECI+), avoids this limitation by using a continuous definition. ECI+ defines the complexity of an economy as the total exports of a country corrected by how difficult it is to export each product and by the size of that country's export economy.

To get the intuition behind ECI+, compare the exports of large aircrafts to that of men's trousers. Large aircraft (unladen weight > 15,000 [kgs]) was the 8th most traded product in 2015 (out of 4,857 products, with USD 157B in total exports)[1], but only three countries (U.S. (37%), France (29%), and Germany (20%)) accounted for more than 86% of total large aircraft exports. Men's trousers, on the other hand, were the 84th most traded product (out of 4,857 products, with USD 24.9B in total exports)[2], but were exported by many countries, including China (22%), Bangladesh (20%), Mexico (5.5%), Pakistan (5.3%), Turkey (5.3%), Germany (3.9%), Italy (3.5%), Vietnam (3.1%), and Tunisia (2.7%). This suggests that exporting one dollar of large aircrafts is, on average, harder than exporting one dollar of trousers; because despite its large export volume, is rare for large aircrafts to represent a substantial fraction of a country's export basket. In this paper, we mathematically formalize this intuition to create an improved measure of economic complexity that estimates the total exports of a country corrected by how difficult it is to export each product and by the size of its export economy.

But, how do we know if we have a good measure of economic complexity? If the goal of a metric of economic complexity is to predict the income generating potential of an economy, then the best measures of economic complexity should be the one that is best at predicting long-term future economic growth. Here, we use this pragmatic criterion to compare our new measure of economic complexity (ECI+) with the original Economic Complexity Index (ECI) [11], and the Fitness Complexity Index [17]. We find that ECI+ outperforms both the original Economic Complexity Index and the Fitness Complexity Index at both short and long-time scales, showing that it is the best measure of economic complexity available. Moreover, we find that ECI+ provides consistent estimators for a wide variety of econometric specifications (OLS, Random Effects, and Fixed Effects models), whereas Fitness Complexity provides inconsistent estimators and is not always significant. Finally, we use our results to predict the average expected annualized growth rates for the next 20 years. These results improve the metrics available to estimate the sophistication of an economy using exports data.

**Data**

We use international trade data from MIT's Observatory of Economic Complexity (atlas.media.mit.edu/about/data[25]). We choose the SITC-4 rev 2 dataset, which provides the longest time series; 1962 to 2014. The dataset captures trade information for 250 countries and 986 products. For the shorter time series, we use the HS92 (4-digit level) BACI dataset, which provides more detailed and precise export data for 226 countries/regions and 1241 products from 1995-2015. To reduce noise, we filter the data by removing city-sized national economies, or economies for which no reliable data was available. That is, we focus on countries with a population of more than 1.25 million in 2008 and exports of more than 1 billion in that year. We also exclude Chad (TCD), Iraq (IRQ), and Afghanistan (AFG). Moreover, we run four time dependent filters. For each year, we exclude products when the dollar value of exports is equal to zero for more than 80% of the countries. In 2010, those products are only 'Copra', 'Manila Hemp', and 'Uranium and Thorium'. We also exclude a country if it's dollar value equals zero for 95% of the products (in 2010 no country would have been excluded). We also exclude a product if global exports are less than 10 million and round to zero any country-product combination that involves less than USD 5,000 in exports. After these filters, our final sample (see SM) for 2010, consists of 121 countries who add to 96.75% of global GDP and 83.37% of global trade.

We use GDP, population, human capital, number of workers, and capital data from the Penn World Tables (PWT 9.0). GDP data is real GDP National Accounts, which measures GDP in constant USD in 2005 [26]. We use measures of institutions including rule of law, voice and accountability, control of corruption, regulatory quality, government effectiveness, political stability, and absence of violence/terrorism, from the World Bank Governance Indicators (2011).

**Methods**

Here we describe the methods used to calculate past metrics of economic complexity (ECI and Fitness), and also, describe the method we introduce to calculate ECI+. We estimate all metrics using the same exact data. In all cases, we let $X_{cp}$ be a matrix summarizing the dollar exports of country $c$ in product $p$.

To estimate the original Economic Complexity Index (ECI), and Fitness Complexity (F), we first need to define a matrix of revealed comparative advantage ($R_{cp}$). This matrix tells us which countries are significant exporters of which products. $R_{cp}$ connects countries to the products they export more than what we expect based on a country's total exports and a product's global market.

Formally, we define:

$$R_{cp} = X_{cp} \Big/ E(X_{cp}) \qquad (1)$$

where $E(X_{cp}) = (\sum_c X_{cp} \sum_p X_{cp})/\sum_{cp} X_{cp}$ is the expected exports of country $c$ in product $p$. This is equal to the size of a country's export economy ($\sum_c X_{cp}$) times the size of that product's global market ($\sum_p X_{cp}$) divided by total world trade ($\sum_{cp} X_{cp}$).

Then, we define $M_{cp}=1$ if a country has $R_{cp} \geq 1$ in a product and $M_{cp}=0$ otherwise. $M_{cp}$ contains information about a country's significant exports. Using $M_{cp}$ we define the diversity of a country ($k_c = \sum_p M_{cp}$), as the number of products that it exports with revealed comparative advantage ($R_{cp} \geq 1$), and the ubiquity of a product ($k_p = \sum_c M_{cp}$), as the number of countries that export that product with revealed comparative advantage ($R_{cp} \geq 1$).

Using the following definitions, the economic complexity index (ECI) and the product complexity index (PCI) are defined by assuming that the complexity of an economy is the average complexity of the products it exports, and the complexity of a product is the average complexity of the countries exporting it. This circular argument gives rise to the following iterative mapping:

$$ECI_c = \frac{1}{k_c} \sum_p M_{cp} PCI_p \qquad (2)$$

$$PCI_p = \frac{1}{k_p} \sum_c M_{cp} ECI_c \qquad (3)$$

Putting (3) into (2) provides an eigenvalue equation whose solution is a country's economic complexity index.

$$ECI_c = \sum_p \frac{M_{cp}}{k_p k_c} \sum_{c'} M_{c'p} ECI_{c'} \qquad (4)$$

The solution for the product complexity index can be obtained by using (4) on (3).

Similarly, the Fitness Complexity index of a country ($F_c$) and of a product ($Q_p$) are defined using a modified version of equations (2) and (3). The Fitness and the associated product complexity are defined as the steady state of the mapping:

$$\tilde{F}_{c,N} = \sum_p M_{cp} Q_{p,N-1} \qquad (4)$$

$$\tilde{Q}_{p,N} = \frac{1}{\sum_c M_{cp} \frac{1}{F_{c,N-1}}} \qquad (5)$$

normalized by its mean after each step:

$$F_{c,N} = \frac{\tilde{F}_{c,N}}{\frac{1}{\{C\}} \sum_c \tilde{F}_{c,N}} \qquad (6)$$

$$Q_{p,N} = \frac{\tilde{Q}_{p,N}}{\frac{1}{\{P\}} \sum_p \tilde{Q}_{p,N}} \qquad (7)$$

where $\{C\}$ and $\{P\}$ are the number of countries and the number of products in the sample respectively, and the initial conditions are $\tilde{F}_{c,o} = 1 \forall c$ and $\tilde{Q}_{p,o} = 1 \forall p$.

Next, we present the formula for ECI+. the metric of economic complexity we introduce in this paper. ECI+ measures the total exports of an economy corrected by how difficult it is to export each product. The intuition is that fewer countries will be able to export products that are more knowledge intensive, even when these products have large markets. So, we correct the dollar export of each product by how "difficult" it is to export it. To calculate this corrected measure of exports we let $X_c^0 = \sum_p X_{cp}$ be the total exports of a country. Also, we let $1/\sum_c \frac{X_{cp}}{X_c^0}$ (one over the average share that a product represents in the average country) be a measure of how difficult it is to export a product. This simply assumes that products that are hard to export will represent a small share of exports for most countries (even when their export volumes are large). Then, we can define the corrected total exports of a country as:

$$X_c^1 = \sum_p \frac{X_{cp}}{\sum_c \frac{X_{cp}}{X_c^0}} \tag{8}$$

$X_c^1$ is a measure of the total exports of a country corrected by how difficult it is to export each product. Yet, we can take this corrected value of total exports to calculate again the share that a product represents of the average country ($X_c^1 \to X_c^2$). This provides us with a second order correction:

$$X_c^2 = \sum_p \frac{X_{cp}}{\sum_c \frac{X_{cp}}{X_c^1}} \tag{9}$$

Taking this intuition to the limit gives us the iterative map:

$$X_c^N = \sum_p \frac{X_{cp}}{\sum_c \frac{X_{cp}}{X_c^{N-1}}} \tag{10}$$

Using this definition we estimate ECI+ as the total exports of a country corrected by how difficult it is to export each product, minus the average share that the country represents in the export of a product (which accounts for the size of a country's export economy).

$$ECI_c^+ = \log(X_c^\infty) - \log\left(\sum_p \frac{X_{cp}}{X_p}\right) \tag{11}$$

To guarantee the numerical convergence of the mapping we normalize $X_c$ at each step (including $X_c^0$) by its geometric mean:

$$X_c^N = \frac{X_c^N}{\left(\prod_{c\prime} X_{c\prime}^N\right)^{\frac{1}{\{C\}}}} \tag{12}$$

where $\{C\}$ is the number of countries in the sample.

Similarly, we define the new product complexity index, PCI+, as the iterations of the mapping:

$$X_p^N = \sum_c \frac{X_{cp}}{\sum_p \frac{X_{cp}}{X_p^{N-1}}} \tag{13}$$

with the initial condition $X_p^0$ being the average share of a product in a country:

$$X_p^0 = \sum_c \frac{X_{cp}}{X_c^0} \tag{14}$$

and also, normalizing at each step (including $X_p^0$) by its geometric mean:

$$X_p^N = \frac{X_p^N}{\left(\prod_{p'} X_{p'}^N\right)^{\frac{1}{\{P\}}}} \tag{15}$$

where $\{P\}$ is the number of products in the sample. Thus, we define $PCI_p^+$ as:

$$PCI_p^+ = \log(X_p) - \log(X_p^\infty) \tag{16}$$

where $X_p$ is the total world trade in a product.

ECI+ and PCI+ are, respectively, a measure of the total exports of a country, corrected by how difficult it is to export each product, and a measure of the total trade in a product, corrected by how easy it is to export that product. Unlike ECI and Fitness, ECI+ has the advantage of not requiring us to discretize the data (as in $M_{cp}$), since it is a function only of export values ($X_{cp}$).

**Results**

We begin by graphically comparing ECI+, with the original Economic Complexity Index (ECI), and Fitness Complexity (F). Then we introduce three econometric models to test the ability of each of these variables to predict future economic growth after controlling for measures of physical capital, human capital, and institutions (OLS, Random Effects, and Fixed Effects). We finalize the results section by looking at the income on future economic growth.

Figure 1 a-c compares the three metrics graphically using a scatter plot for all countries. As expected, the metrics are positively correlated but have important deviations among them. ECI+ and ECI have a strong correlation ($R^2=85\%$), but ECI+ tends to rank manufacturing heavy countries higher than ECI (ECI+ ranks Vietnam higher than Qatar and China higher than Norway). On the other hand, the correlation between Fitness, and both, ECI and ECI+, is much lower (respectively 48% and 43%). Fitness Complexity ranks many Southern European countries (such as Spain, Italy, and Portugal) at the top of the ranking, and also, provides very low complexity values for advanced East Asian and European economics, such as South Korea, Switzerland, Finland, Japan, and Singapore.

Next, we compare the three complexity measures with GDP per capita (Figure 1 d-f). As expected, all metrics show a positive correlation with income. Yet, Fitness Complexity has a relatively weak correlation with income levels compared to both ECI+ and ECI.

Figure 1 g-i tries to unpack the difference between these three metrics by comparing them with a pure measure of diversity (number of products a country exports with $R_{cp}>1$). Unlike ECI+ and ECI, Fitness Complexity tracks the raw diversity of countries closely, suggesting that it is not sensitive to differences in the sophistication of products and does not provide much additional information than using a simple measure of diversity. This explains why, for the Fitness measure, the economy of Greece is ranked higher than that of Japan, Sweden, or China.

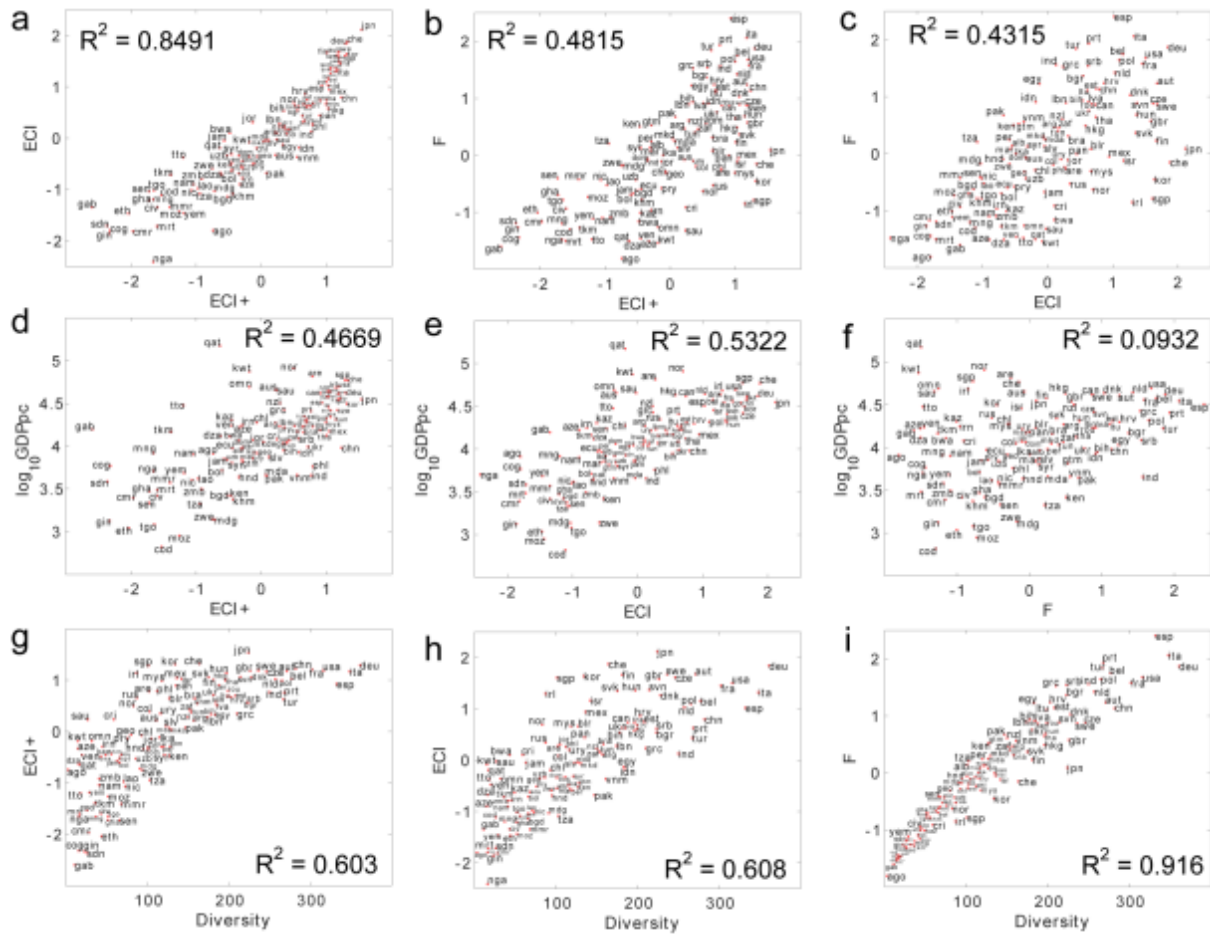

*Figure* 1 *Comparison of the three measures of economic complexity (ECI+, ECI, and F) using 2010 data. (**a-c**) Correlation between each pair of measures: **a** ECI and ECI+, **b** F and ECI+, and **c** F and ECI. (d-f) Correlation between each of the three measures with GDP per capita. (g-i) Correlation between each of the three measures with a pure measure of diversity (number of products a country exports with Rcp>1).*

Next, we compare the ability of these three measures to predict future economic growth using the most consistent country set and data we have available (1973 – 2013) (see SM for details). We start with a basic 40-year cross section model, and then specify 5, 10, and 20 year average periods using pooled OLS, random effects, and fixed effects models. The pooled OLS model provides a baseline, but suffers from omitted variable bias. Random effects models allow us to breakdown the variance into that explained by between and within country variation. Fixed effects regressions help us account for some of the omitted variable bias due to time or individual consistent effects and to relax the assumption that the country specific effects are uncorrelated with our explanatory factors (which is problematic for Random Effects models).

In addition to the long run dataset, we use the three models with five-year panels from 1998 – 2014 to include data from the World Bank Governance Indicators (which is only available for this period). For these specifications, we use exports data from the BACI dataset compiled by CEPII [27]. This exports data starts in 1996 and is aggregated according to the HS4 classification.

We start by comparing the performance of these three metrics for the entire period for which data is available. Table 1 uses a cross sectional OLS regression to predict annualized growth from 1973 – 2013. Our variables of interest are ECI+, ECI, and Fitness, holding constant initial levels of income, human capital, capital per worker, and population. The regressions provide a broad picture of long-term economic growth. Though we can't rely on the OLS for the coefficient estimates, we still find that ECI+ and ECI, along with their interaction terms with GDP per capita, are significant at the 1% level while Fitness (F) is not significant at any of the conventional levels. As expected, we also find that initial income is negatively correlated with growth in all three regressions and that human capital is positive and significant. From our variables of interest, ECI+ explains most of the variance in growth, although it is only marginally better than ECI. Fitness is a far third. In fact, ECI+ and ECI explain more than 8 percentage points of variance more than F.

*Table 1 Linear regression model to predict economic growth using each of the three measures of economic complexity.*

Cross-Sectional 40 Year Growth (1973- 2013)

|  | (1) | (2) | (3) |
|---|---|---|---|
| Initial ECI+ | 0.0416*** <br> (0.013) |  |  |
| Ini GDPpc * ECI+ | -0.00395*** <br> (0.001) |  |  |
| Initial ECI |  | 0.0462*** <br> (0.016) |  |
| Ini GDPpc * ECI |  | -0.00440*** <br> (0.002) |  |
| Initial F |  |  | 0.0132 <br> (0.016) |
| Ini GDPpc * F |  |  | -0.00119 <br> (0.002) |
| Initial GDPpc | -0.0111*** <br> (0.003) | -0.0114*** <br> (0.004) | -0.00955** <br> (0.004) |
| Initial Human Capital | 0.00984*** <br> (0.004) | 0.00996*** <br> (0.003) | 0.0125*** <br> (0.004) |
| Initial Pop | 0.0000128* <br> (0.000) | 0.0000156* <br> (0.000) | 0.0000201 <br> (0.000) |
| Initial Capital | -0.000259 <br> (0.003) | -0.000410 <br> (0.003) | 0.000342 <br> (0.003) |
| Constant | 0.101*** <br> (0.014) | 0.106*** <br> (0.014) | 0.0739*** <br> (0.014) |
| N | 87 | 87 | 87 |
| $\bar{R}^2$ adjusted | 0.485 | 0.479 | 0.396 |
| RMSE | 0.0114 | 0.0115 | 0.0124 |
| N clusters | 87 | 87 | 87 |

Standard errors in parentheses
Robust-clustered standard errors
* $p < 0.10$, ** $p < 0.05$, *** $p < 0.01$

Next, we present the results of predictions for 5, 10 and 20-year intervals, using OLS panels, random effects, and fixed effect models (Table 2-Table **4**). These intervals help smooth over fluctuations in growth rates that may occur from a variety of shocks, such as natural disasters or natural resource price volatility. We balance our panel data set and only include countries for which we have information for the entire 1973–2013 period. This is to ensure that countries that drop in and out of the data are not driving our estimates.

We compare estimates of complexity measures with pooled OLS, random effects, and fixed effects regressions using the base model:

$$Growth_j = \beta_1 GDP_{j,t-5} + \beta_2 ECI_{i,j,t-5} + \beta_3 GDP_{i,j,t-5} ECI_{i,j,t-5} + \gamma C'_{j,t-5} + \delta D_t + v_{i,t}$$

Where *Growth<sub>j</sub>* is the compound annualized growth rate for country *j* during the period observed (CAGR). *GDP* is the initial GDP per capita of country *j* at the beginning of the period (time *t* minus 5 years in this example, but 10 or 20 years in other regressions), and ECI stands for ECI+, ECI, or Fitness, depending on the model. *C* is a vector of control variables that include initial population levels, initial human capital, and initial capital per worker. *D<sub>t</sub>* are dummy variables for the respective year to control for any time effects on growth, such as a global economic recession. The error term is $v_{i,t}$, in the case of the random effects model, this includes the individual effect coefficients, as well. Our model includes an interaction term between GDP per capita and complexity because this term helps test whether the effect of complexity on growth is more important for lower or higher income countries, and also, because this term has been shown to be a significant predictor in previous studies.[22]

We present the results of the OLS, random effects, and fixed effects models for the 5, 10, and 20-year intervals in Table 2-Table 4. There are several features that are common to all specifications. First, we note that the coefficients for ECI+ and its interaction with income are consistent and statistically significant across all specifications. The coefficients of ECI behave similarly, with the exception of the five-year fixed effect model, where the coefficient preserves its positive sign but is not statistically significant. The coefficients for Fitness, however, are inconsistent and not significant in the fixed effect models (and even reverse signs). Second, we find in the pooled OLS that the adjusted-$R^2$ of ECI+ is always larger than that of ECI and Fitness. Third, the random effects model reveals that most of the variation is between countries. Moreover, we find that the coefficient for initial human capital is generally positive and significant and that of initial income is generally negative and significant.

*Table 2* OLS, Random Effects, and Fixed Effects 5-year intervals linear growth regression models.

Annualized 5 Year Growth Rates

|  | ECI+ | | | ECI | | | F | | |
|---|---|---|---|---|---|---|---|---|---|
|  | (1) OLS | (2) RE | (3) FE | (4) OLS | (5) RE | (6) FE | (7) OLS | (8) RE | (9) FE |
| Initial ECI+ | 0.0563*** (0.012) | 0.0570*** (0.013) | 0.0621** (0.026) | | | | | | |
| Ini GDPpc * Ini ECI+ | -0.00508*** (0.001) | -0.00533*** (0.001) | -0.00720** (0.003) | | | | | | |
| Initial ECI | | | | 0.0679*** (0.015) | 0.0612*** (0.016) | 0.0249 (0.016) | | | |
| Ini GDPpc * Ini ECI | | | | -0.00633*** (0.002) | -0.00569*** (0.002) | -0.00268 (0.002) | | | |
| Initial F | | | | | | | 0.0457*** (0.014) | 0.0381** (0.016) | -0.00000393 (0.037) |
| Ini GPCpc * F | | | | | | | -0.00433*** (0.001) | -0.00352** (0.002) | 0.000552 (0.004) |
| Initial GDPpc | -0.0109*** (0.004) | -0.0133*** (0.004) | -0.0443*** (0.009) | -0.00931** (0.004) | -0.0120*** (0.004) | -0.0424*** (0.010) | -0.0101** (0.005) | -0.0128*** (0.005) | -0.0408*** (0.010) |
| Initial Human Capital | 0.0100*** (0.004) | 0.0133*** (0.004) | -0.00318 (0.016) | 0.00940** (0.004) | 0.0117*** (0.004) | -0.00486 (0.015) | 0.0136*** (0.004) | 0.0156*** (0.004) | -0.00625 (0.015) |
| Initial Pop | 0.000296 (0.001) | 0.000697 (0.001) | 0.0156 (0.011) | 0.00153 (0.001) | 0.00169 (0.001) | 0.0156 (0.011) | 0.000865 (0.001) | 0.00103 (0.001) | 0.0160 (0.011) |
| Initial Capital | -0.00278 (0.003) | -0.00187 (0.003) | 0.00311 (0.008) | -0.00314 (0.003) | -0.00215 (0.003) | 0.00126 (0.008) | -0.000281 (0.003) | 0.000385 (0.003) | 0.000849 (0.008) |
| Constant | 0.128*** (0.015) | 0.132*** (0.018) | 0.334*** (0.102) | 0.116*** (0.016) | 0.123*** (0.019) | 0.368*** (0.094) | 0.0833*** (0.015) | 0.0913*** (0.017) | 0.357*** (0.099) |
| N | 716 | 716 | 716 | 716 | 716 | 716 | 716 | 716 | 716 |
| $\bar{R}^2$ within | | 0.200 | 0.301 | | 0.201 | 0.292 | | 0.198 | 0.294 |
| $\bar{R}^2$ between | | 0.409 | 0.0349 | | 0.361 | 0.0372 | | 0.305 | 0.0483 |
| $\bar{R}^2$ overall | | 0.252 | 0.0420 | | 0.240 | 0.0419 | | 0.220 | 0.0474 |
| $\bar{R}^2$ adjusted | 0.242 | | 0.288 | 0.229 | | 0.279 | 0.211 | | 0.281 |
| RMSE | 0.0262 | 0.0248 | 0.0218 | 0.0264 | 0.0248 | 0.0220 | 0.0267 | 0.0249 | 0.0219 |
| N_clust | 91 | 91 | 91 | 91 | 91 | 91 | 91 | 91 | 91 |

Standard errors in parentheses
Robust-clustered standard errors & Year FE
* $p < 0.10$, ** $p < 0.05$, *** $p < 0.01$

*Table 3* OLS, Random Effects, and Fixed Effects 10-year intervals linear growth regression models.

Annualized 10 Year Growth Rates

|  | ECI+ | | | ECI | | | F | | |
|---|---|---|---|---|---|---|---|---|---|
|  | (1) OLS | (2) RE | (3) FE | (4) OLS | (5) RE | (6) FE | (7) OLS | (8) RE | (9) FE |
| Initial ECI+ | 0.0527*** (0.012) | 0.0505*** (0.011) | 0.0385* (0.021) | | | | | | |
| Ini GDPpc * Ini ECI+ | -0.00458*** (0.001) | -0.00442*** (0.001) | -0.00404* (0.002) | | | | | | |
| Initial ECI | | | | 0.0711*** (0.016) | 0.0660*** (0.016) | 0.0304** (0.013) | | | |
| Ini GDPpc * Ini ECI | | | | -0.00671*** (0.002) | -0.00625*** (0.002) | -0.00356** (0.002) | | | |
| Initial F | | | | | | | 0.0397*** (0.013) | 0.0314** (0.015) | -0.0122 (0.036) |
| Ini GPCpc *Ini F | | | | | | | -0.00373*** (0.001) | -0.00287* (0.002) | 0.00202 (0.004) |
| Initial GDPpc | -0.0126*** (0.004) | -0.0150*** (0.004) | -0.0502*** (0.009) | -0.0111*** (0.004) | -0.0137*** (0.004) | -0.0495*** (0.010) | -0.0110** (0.004) | -0.0141*** (0.005) | -0.0467*** (0.010) |
| Initial Human Capital | 0.00936** (0.004) | 0.0117*** (0.004) | -0.00106 (0.016) | 0.0104*** (0.004) | 0.0128*** (0.004) | 0.00253 (0.016) | 0.0141*** (0.004) | 0.0164*** (0.004) | -0.000850 (0.016) |
| Initial Pop | -0.0000522 (0.001) | 0.000123 (0.001) | 0.0136 (0.011) | 0.00146 (0.001) | 0.00157 (0.001) | 0.0136 (0.011) | 0.000847 (0.001) | 0.000970 (0.001) | 0.0134 (0.011) |
| Initial Capital | -0.00184 (0.003) | -0.000822 (0.003) | 0.00342 (0.008) | -0.00216 (0.003) | -0.00103 (0.003) | 0.00359 (0.008) | 0.000244 (0.004) | 0.00120 (0.003) | 0.00206 (0.008) |
| Constant | 0.144*** (0.017) | 0.149*** (0.018) | 0.386*** (0.089) | 0.129*** (0.016) | 0.134*** (0.016) | 0.372*** (0.083) | 0.0887*** (0.015) | 0.101*** (0.017) | 0.367*** (0.085) |
| N | 357 | 357 | 357 | 357 | 357 | 357 | 357 | 357 | 357 |
| $\bar{R}^2$ within | | 0.250 | 0.445 | | 0.256 | 0.442 | | 0.252 | 0.447 |
| $\bar{R}^2$ between | | 0.426 | 0.0543 | | 0.382 | 0.0443 | | 0.289 | 0.0616 |
| $\bar{R}^2$ overall | | 0.311 | 0.0645 | | 0.297 | 0.0576 | | 0.251 | 0.0694 |
| $\bar{R}^2$ adjusted | 0.296 | | 0.431 | 0.282 | | 0.427 | 0.239 | | 0.433 |
| RMSE | 0.0211 | 0.0197 | 0.0150 | 0.0214 | 0.0197 | 0.0151 | 0.0220 | 0.0198 | 0.0150 |
| N_clust | 91 | 91 | 91 | 91 | 91 | 91 | 91 | 91 | 91 |

Standard errors in parentheses
Robust-clustered standard errors & Year FE
* $p < 0.10$, ** $p < 0.05$, *** $p < 0.01$

*Table 4 OLS, Random Effects, and Fixed Effects 20-year intervals linear growth regression models.*

Annualized 20 Year Growth Rates

|  | ECI+ | | | ECI | | | F | | |
|---|---|---|---|---|---|---|---|---|---|
|  | (1) OLS | (2) RE | (3) FE | (4) OLS | (5) RE | (6) FE | (7) OLS | (8) RE | (9) FE |
|  | (1) | (2) | (3) | (4) | (5) | (6) | (7) | (8) | (9) |
| Initial ECI+ | 0.0482*** (0.011) | 0.0456*** (0.011) | 0.0457** (0.021) | | | | | | |
| Ini GPCpc * Ini ECI+ | -0.00419*** (0.001) | -0.00399*** (0.001) | -0.00494** (0.002) | | | | | | |
| Initial ECI | | | | 0.0614*** (0.015) | 0.0606*** (0.016) | 0.0603** (0.024) | | | |
| Ini GPCpc * Ini ECI | | | | -0.00581*** (0.002) | -0.00577*** (0.002) | -0.00713*** (0.003) | | | |
| Initial F | | | | | | | 0.0270** (0.012) | 0.0234* (0.014) | 0.0126 (0.043) |
| Ini GPCpc * Ini F | | | | | | | -0.00252* (0.001) | -0.00221 (0.001) | -0.00102 (0.005) |
| Initial GDPpc | -0.0120*** (0.003) | -0.0134*** (0.003) | -0.0418*** (0.007) | -0.0114*** (0.004) | -0.0136*** (0.004) | -0.0444*** (0.007) | -0.0103** (0.004) | -0.0125*** (0.004) | -0.0405*** (0.008) |
| Initial Human Capital | 0.00782** (0.003) | 0.00888*** (0.003) | -0.0118 (0.017) | 0.00976*** (0.003) | 0.0112*** (0.003) | -0.00570 (0.015) | 0.0134*** (0.004) | 0.0147*** (0.004) | -0.0129 (0.017) |
| Initial Pop | 0.000144 (0.001) | 0.000274 (0.001) | 0.00102 (0.009) | 0.00164* (0.001) | 0.00172* (0.001) | 0.00119 (0.008) | 0.00132 (0.001) | 0.00152 (0.001) | -0.00130 (0.010) |
| Initial Capital | -0.00150 (0.003) | -0.000731 (0.003) | 0.00111 (0.005) | -0.00158 (0.003) | -0.000576 (0.003) | 0.00279 (0.005) | 0.000255 (0.003) | 0.00101 (0.003) | 0.000416 (0.006) |
| Constant | 0.133*** (0.017) | 0.134*** (0.018) | 0.415*** (0.073) | 0.122*** (0.016) | 0.127*** (0.017) | 0.408*** (0.067) | 0.0807*** (0.015) | 0.0879*** (0.017) | 0.416*** (0.069) |
| N | 182 | 182 | 182 | 182 | 182 | 182 | 182 | 182 | 182 |
| $\bar{R}^2$ within | | 0.372 | 0.659 | | 0.416 | 0.673 | | 0.381 | 0.650 |
| $\bar{R}^2$ between | | 0.477 | 0.0655 | | 0.415 | 0.0469 | | 0.331 | 0.0582 |
| $\bar{R}^2$ overall | | 0.436 | 0.0894 | | 0.415 | 0.0728 | | 0.349 | 0.0821 |
| $\bar{R}^2$ adjusted | 0.416 | | 0.645 | 0.394 | | 0.660 | 0.328 | | 0.636 |
| RMSE | 0.0160 | 0.0142 | 0.00816 | 0.0163 | 0.0138 | 0.00798 | 0.0172 | 0.0142 | 0.00826 |
| N_clust | 95 | 95 | 95 | 95 | 95 | 95 | 95 | 95 | 95 |

Standard errors in parentheses
Robust-clustered standard errors & Year FE

Next, we control for institutions by adding data from the World Bank's Governance Indicators in five-year panels. Table 5 repeats the OLS, random effects, and fixed effects models using 5-year panels. We observe that the coefficients of ECI+ are consistent across the three models after we control for institutions. The effects of initial income per capita and human capital are also unchanged. Among the institutional variables regulatory quality is consistently negative and significant, political stability and control of corruption are positive and significant, government effectiveness is consistently positive but not significant, and initial law and voice and accountability are inconsistent.

*Table 5* Growth regression controlling for institutions.

Annualized 5 Year Growth Rates

|  | ECI+ | | | ECI | | | F | | |
|---|---|---|---|---|---|---|---|---|---|
|  | (1) OLS | (2) RE | (3) FE | (4) OLS | (5) RE | (6) FE | (7) OLS | (8) RE | (9) FE |
| Initial ECI+ | 0.0662*** (0.019) | 0.0726*** (0.017) | 0.0828* (0.043) | | | | | | |
| Ini GDPpc* Ini ECI+ | -0.00661*** (0.002) | -0.00726*** (0.002) | -0.0108** (0.005) | | | | | | |
| Initial ECI | | | | 0.0877*** (0.024) | 0.100*** (0.025) | 0.0975 (0.061) | | | |
| Ini GDPpc* Ini ECI | | | | -0.00901*** (0.002) | -0.0102*** (0.002) | -0.0103* (0.006) | | | |
| Initial F | | | | | | | 0.0567*** (0.017) | 0.0699*** (0.017) | 0.0628 (0.079) |
| Ini GDPpc * Ini F | | | | | | | -0.00569*** (0.002) | -0.00700*** (0.002) | -0.00658 (0.008) |
| Initial GDPpc | -0.0186*** (0.004) | -0.0226*** (0.005) | -0.0729*** (0.023) | -0.0175*** (0.005) | -0.0217*** (0.005) | -0.0751*** (0.023) | -0.0194*** (0.005) | -0.0243*** (0.005) | -0.0768*** (0.023) |
| Law | -0.00299 (0.006) | -0.00886 (0.006) | -0.0284*** (0.009) | -0.00193 (0.006) | -0.00658 (0.006) | -0.0240** (0.011) | -0.00910 (0.006) | -0.0156*** (0.006) | -0.0307*** (0.009) |
| Voice and accountability | -0.00814** (0.004) | -0.00611 (0.004) | 0.0115* (0.006) | -0.00930** (0.004) | -0.00750* (0.004) | 0.0119* (0.006) | -0.00749** (0.004) | -0.00520 (0.004) | 0.0130** (0.006) |
| Control of corruption | 0.00942 (0.007) | 0.0182*** (0.006) | 0.0138* (0.007) | 0.00876 (0.007) | 0.0175*** (0.006) | 0.0128 (0.008) | 0.00789 (0.007) | 0.0175*** (0.006) | 0.0148* (0.008) |
| Regulatory Quality | -0.0106** (0.005) | -0.0172*** (0.005) | -0.0247*** (0.007) | -0.0105* (0.006) | -0.0174*** (0.006) | -0.0275*** (0.007) | -0.00718 (0.006) | -0.0147** (0.006) | -0.0263*** (0.007) |
| Government effectiveness | 0.00220 (0.008) | 0.00243 (0.007) | 0.0224** (0.010) | 0.00364 (0.008) | 0.00276 (0.007) | 0.0203** (0.010) | 0.00353 (0.007) | 0.00359 (0.007) | 0.0198* (0.010) |
| Political Stability | 0.00509** (0.002) | 0.00455* (0.003) | -0.000737 (0.006) | 0.00546** (0.002) | 0.00499** (0.002) | 0.000606 (0.005) | 0.00515** (0.002) | 0.00476* (0.003) | 0.000522 (0.005) |
| Human Capital | 0.0217*** (0.004) | 0.0242*** (0.004) | 0.0395 (0.025) | 0.0236*** (0.004) | 0.0256*** (0.004) | 0.0271 (0.024) | 0.0259*** (0.004) | 0.0289*** (0.004) | 0.0229 (0.023) |
| Pop | 0.000157 (0.001) | 0.000168 (0.002) | 0.0656*** (0.025) | 0.00122 (0.001) | 0.00121 (0.001) | 0.0582** (0.027) | 0.000940 (0.001) | 0.000889 (0.001) | 0.0443* (0.023) |
| Capital | 0.0000997 (0.004) | 0.00282 (0.004) | 0.0147 (0.019) | -0.000984 (0.004) | 0.00157 (0.004) | 0.0108 (0.016) | 0.00391 (0.005) | 0.00747 (0.005) | 0.0141 (0.016) |
| Constant | 0.140*** (0.032) | 0.142*** (0.033) | 0.265 (0.288) | 0.143*** (0.036) | 0.145*** (0.041) | 0.376 (0.249) | 0.0871*** (0.032) | 0.0865** (0.036) | 0.412* (0.235) |
| N | 273 | 273 | 273 | 273 | 273 | 273 | 273 | 273 | 273 |
| $\bar{R}^2$ within | | 0.433 | 0.577 | | 0.445 | 0.573 | | 0.454 | 0.564 |
| $\bar{R}^2$ between | | 0.320 | 0.0998 | | 0.321 | 0.107 | | 0.276 | 0.111 |
| $\bar{R}^2$ overall | | 0.367 | 0.0788 | | 0.373 | 0.0844 | | 0.352 | 0.0918 |
| $\bar{R}^2$ adjusted | 0.346 | | 0.555 | 0.351 | | 0.550 | 0.331 | | 0.540 |
| RMSE | 0.0208 | 0.0177 | 0.0126 | 0.0207 | 0.0176 | 0.0127 | 0.0211 | 0.0177 | 0.0128 |
| N Countries | 91 | 91 | 91 | 91 | 91 | 91 | 91 | 91 | 91 |

Standard errors in parentheses
Robust-clustered standard errors & Year FE
* $p < 0.10$, ** $p < 0.05$, *** $p < 0.01$

Finally, we use our twenty-year OLS model (Table 4, Columns 1, 4, and 7) to provide predictions of annualized future economic growth for the period between 2013 and 2033. Figure 2 shows maps for the predictions obtained for ECI+, ECI, and F. We note that the OLS predictions have a strong regression to the mean, so the actual values should not be as informative as the relative rankings. Nevertheless, the maps still show some interesting patterns. All maps coincide in the continued growth of China, India, Philippines, and much of Eastern Europe. Yet, ECI+ is more optimistic than the other two predictors in the future economic growth of Peru, Mexico, and Mongolia, and less optimistic in the growth of East African economies (for complete table see Appendix).

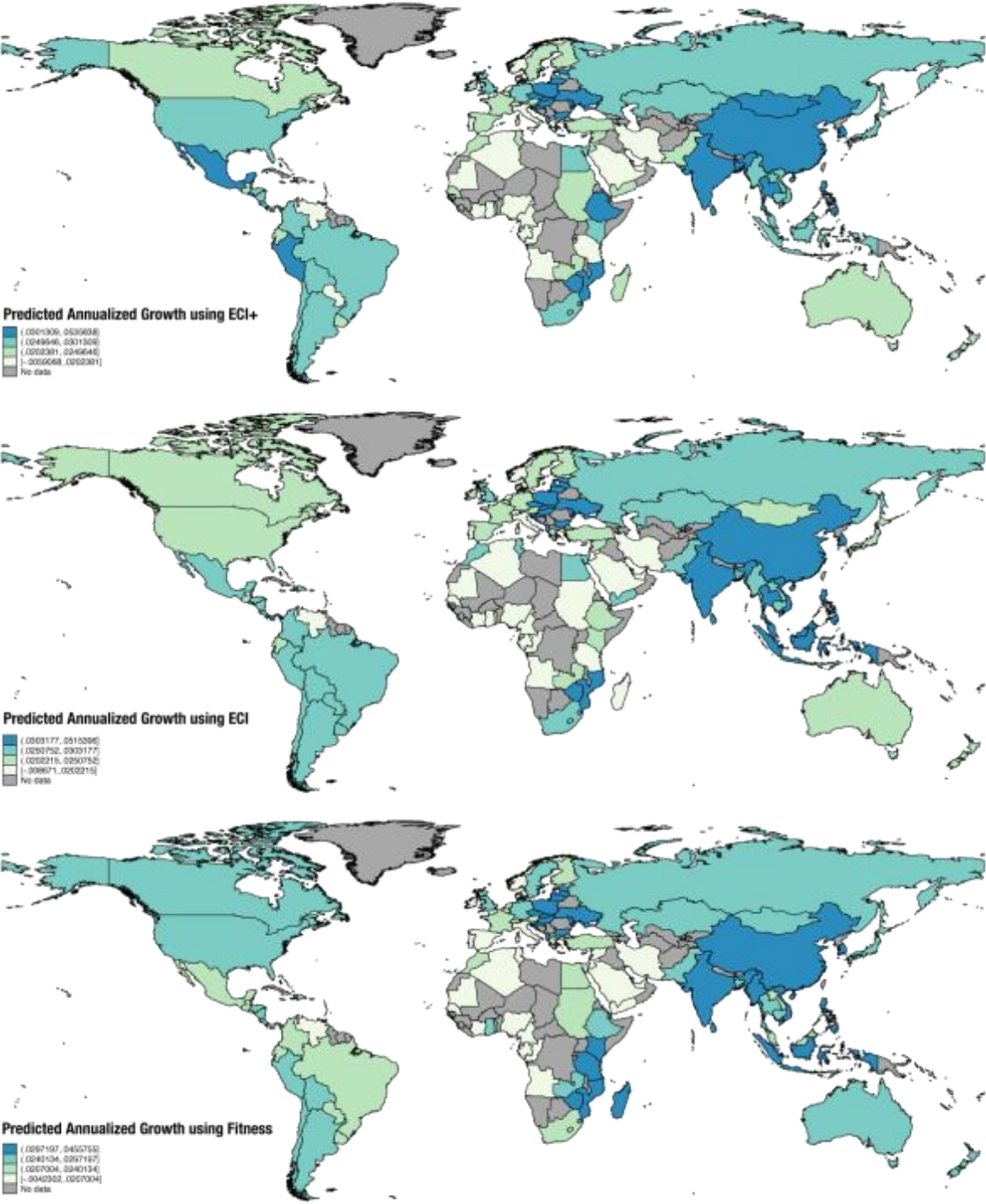

*Figure 2 Maps showing the predictions of annualized future economic growth for the period between 2013 and 2033 for ECI+, ECI, and F.*

**Discussion**

Understanding economic growth is one of the central questions of development economics. In recent years, measures of economic complexity designed to capture the knowledge intensity of economies, have become important indicators of an economy's future economic growth potential. Here we provide a methodological contribution by developing an improved metric of economic complexity that is better at predicting future economic growth than previously proposed measures. The metric is based on a method that corrects the exports of a country by considering how difficult it is to export each product. OLS, random effects, and fixed effects models show that this measure is better at predicting long-term growth than the previously proposed measures and has very consistent values for the estimators across all specifications.

Yet, economic complexity is not the only factor that helps predict future economic growth. At the margin of Economic Complexity, human capital has a mostly consistent positive effect, and income has a consistently negative effect in agreement with the traditional economic growth literature. When it comes to institutions, we find the effects of political stability and control of corruption to be mostly positive and significant, while those of regulatory quality tend to be negative and significant.

This paper contributes to the growing literature on economic complexity by advancing an improved metric to estimate the total knowledge content of an economy and its income generating capacity.

---

[1] Source: MIT's Observatory of Economic Complexity
http://atlas.media.mit.edu/en/profile/hs92/880240/

[2] Source: MIT's Observatory of Economic Complexity
http://atlas.media.mit.edu/en/profile/hs92/620342/